\documentclass[preprint]{revtex4}
\usepackage[american]{babel}
\usepackage{bm}
\usepackage{amsfonts}
\usepackage{graphicx}
\usepackage{amsmath}

\newcommand{\new}{}

\begin{document}

\title{Semiclassical Vlasov and fluid models for an electron gas with spin effects}
\author{J\'{e}r\^{o}me Hurst, Omar Morandi, Giovanni Manfredi, Paul-Antoine Hervieux}
\affiliation{Institut de Physique et Chimie des Mat\'{e}riaux de
Strasbourg and Labex NIE, \\ Universit\'{e} de Strasbourg, CNRS UMR 7504 \\ BP 43 -- F-67034 Strasbourg Cedex 2, France}

\date{\today}

\begin{abstract}
We derive a four-component Vlasov equation for a system composed of spin-1/2 fermions (typically electrons). The orbital part of the motion is classical, whereas the spin degrees of freedom are treated in a completely quantum-mechanical way. The corresponding hydrodynamic equations are derived by taking velocity moments of the phase-space distribution function. This hydrodynamic model is closed using a maximum entropy principle in the case of three or four constraints on the fluid moments, both for Maxwell-Boltzmann and Fermi-Dirac statistics.

\end{abstract}
\maketitle

\section{Introduction}

The coupling between the electronic dynamics and the spin degrees of freedom in nanometric objects has stimulated a great deal of interest, both theoretical and experimental, over the last few decades.
Many experimental studies have concentrated on the
charge dynamics of an electron gas confined in metallic
nanostructures such as thin films
\cite{Suarez,Bigot},
nanotubes \cite{Lauret}, metal clusters \cite{Schlipper,Campbell}
and nanoparticles \cite{Voisin,dionne, pendry}.
From the theoretical point of view, earlier works were based on phenomenological
models \cite{Rethfeld,Aesch00,Guil03}
that employed Boltzmann-type equations within the framework of Fermi-liquid
theory \cite{Pines}. Studies based on microscopic models (either classical or quantum) are more recent and limited to relatively small systems, due to their considerable computational cost.
In the quantum regime, the ultrafast electron
dynamics in metallic clusters and nanopatricles was studied by Calvayrac {\it et al.} \cite{Calvayrac} and more recently Teperik et al. \cite{borisov} using the time-dependent density functional
theory (DFT). The
many-particle quantum dynamics of the electron gas in a thin metal
film was studied by Schwengelbeck {\it et al.} \cite{Sch00} within
the time-dependent Hartree-Fock (HF) approximation.

The semiclassical limit of the above quantum models (DFT and HF) is
 the self-consistent Vlasov-Poisson system. The
Vlasov-Poisson model was used to perform particle-in-cell (PIC)
simulations of the electron dynamics in  metal clusters
\cite{Calvayrac,Dal03}, and to obtain analytical results in the
linear regime for metal clusters \cite{Fom99} and thin films
\cite{Zar04}.
The nonlinear electron response of thin metal films was studied by Manfredi and Hervieux \cite{mh}, who identified a ballistic electronic modes generated by bunches of electrons bouncing back and forth on the film surfaces. These works were later extended to the quantum domain using Wigner transforms \cite{metal_films_3}.

The above studies included the charge, but not the spin degrees of freedom.
However, it is well known that spin effects (particularly the Zeeman splitting and the spin-orbit coupling) can play a decisive role in nanometric systems such as semiconductor quantum dots \cite{puente,serra} and diluted magnetic semiconductors \cite{morandi_NJP_09,morandi_dms_prB}.
Early experiments on magnetic films \cite{Beaurep} showed that the electron spins respond to an external optical excitation on a subpicosecond timescale, which is the typical timescale for the electrons to equilibrate thermally with the lattice in a metallic nanostructure. From a fundamental point of view, several mechanisms have been proposed for the modification of the magnetic order of nanostructures subject to an ultrafast external field, ranging from the spin-orbit coupling \cite{Zhang} to the spin-lattice interactions \cite{Koopmans}. More recent
experiments \cite{Bigot_natphys} have shown the existence of a coherent coupling between a femtosecond laser pulse and the magnetization of a ferromagnetic thin film. A recent review of the state of the art in the field of ultrafast magnetization dynamics in nanostructures can be found in Ref. \cite{Bigot_annphys}.

In the present work, we propose a semiclassical mean-field model, based on the Vlasov equation, which includes the orbital motion in a classical fashion but incorporates spin effects in a fully quantum-mechanical way.
The Vlasov model is derived using the phase-space formulation of quantum mechanics due to Wigner \cite{Wigner}. The spin enters the model via the Zeeman effect (coupling of the spin with a magnetic field, either external or self-consistent), which is the first non-relativistic correction to the spinless dynamics. The spin-orbit coupling is a second-order (in $1/c$) correction that will be neglected here, although it could be included with relative ease in our model. Recent results on this and other relativistic corrections may be found in Refs. \cite{anant,hinsch}.

Subsequently, we will derive the corresponding hydrodynamic (or fluid) equations by taking velocity moments of the Vlasov equation.
Spinless hydrodynamic methods have been successfully used in the past to model the electron dynamics in molecular systems \cite{qhy_mol}, metal clusters and nanoparticles \cite{qhy_clust1,qhy_clust2,giovani_paper}, thin metal films \cite{qhy_thin_met}, quantum plasmas \cite{qhy_plasma1,qhy_plasma2} and semiconductors \cite{qhy_semi_cond}.
Hydrodynamic equations including the spin degrees of freedom were derived by Brodin and Marklund \cite{Brodin} using the Madelung transformation of the wave function \cite{Madelung}. More recently, a relativistic hydrodynamic model was obtained by Asenjo et al. \cite{asenjo} from the Dirac equation. These approaches based on the Madelung transformation usually lead to cumbersome equations that are in practice very hard to solve, either analytically or numerically, even in the nonrelativistic limit. Our technique, which separates clearly the (classical) orbital motion from the (quantum) spin dynamics, leads to a simpler and more transparent fluid model, where the meaning of each term in the equations is more intuitive.

The fluid equations derived from the Vlasov model constitute an infinite hierarchy of equations that need to be closed using some additional physical hypotheses. Although this is relatively easy for spinless systems (where the closure can be obtained by a assuming a suitable equation of state), things are far subtler when the spin degrees of freedom are included. Here, we shall employ a general procedure based on the maximization of entropy. Using this approach, we obtain a closed set of fluid equations for both Maxwell-Boltzmann and the Fermi-Dirac statistics, keeping up to four fluid moments of the Vlasov distribution function.

\section{Derivation of the spin Vlasov model}\label{sec:vlasov}
We consider an ensemble of spin-1/2 particles (electrons) in the presence of a magnetic field $\bm{B}$ and a electric potential $V$. We denote the  Schr\"odinger wave function of the $\mu-$th particle state by
\begin{equation}
\Psi_{\mu}(\bm{r},t) =\Psi_{\mu}^{\uparrow}(\bm{r},t)\, \left|\uparrow \right\rangle +\Psi_{\mu}^{\downarrow}(\bm{r},t)\, \left|\downarrow \right\rangle, \label{pauli-spinor}
\end{equation}
where $ \Psi_{\mu}^{\uparrow}(\bm{r},t)$  and $\Psi_{\mu}^{\downarrow}(\bm{r},t)$ are respectively the spin-up and spin-down components of the wave function,  $\bm{r}$  denotes the spatial position, and $t$ the time. The  evolution of the system is governed by the Pauli-Schr\"odinger equation
\begin{equation}
i\hbar \frac{\partial \Psi_{\mu}(\bm{r},t)}{\partial t} =
\left[ \left( -\frac{\hbar ^{2}}{2m} \bm{\nabla}^{2} + V(\bm{r},t) \right) \sigma_{0} + \mu _{B} \bm{\sigma} \cdot \bm{B}(\bm{r},t) \right] \Psi_{\mu}(\bm{r},t).
\label{ks equation}
\end{equation}
Here, $\mu_{B}=e\hbar/2m$ is the Bohr magneton, $ \bm{ \sigma} $ is the vector of the $2 \times 2$ Pauli matrices, and $\sigma_{0}$ is the $2 \times 2$ identity matrix. In Eq. \eqref{ks equation} the electromagnetic fields can be either external or self-consistently generated by the particle charge density and current.

When the fields are self-consistent, the system composed of Eq. \eqref{ks equation} together with Maxwell's equations (or an appropriate nonrelativistic limit thereof \cite{anant,manfredi_maxwell}) constitute a mean-field approximation to the exact N-body dynamics. This mean-field approach can also be extended, in the spirit of density functional theory (DFT), to include exchange and correlation effects by adding suitable potentials and fields that are functionals of the electron density  \cite{metal_films_2}. The resulting equations are potentially equivalent to the exact N-body treatment, although the exchange-correlations functionals are not known and need to be somehow approximated.

As an alternative to the Schr\"odinger framework, a statistical ensemble of quantum particles is more conveniently described by a density matrix formalism. Here, we will make use of the phase-space formulation of the quantum dynamics due to Wigner \cite{Wigner}, which is equivalent to the density matrix approach and provides the considerable advantage that the equation of motion bears a strong similarity with the classical Vlasov description. Furthermore, in the Wigner formalism, the classical limit can be easily evaluated and the quantum corrections to the Vlasov equation are obtained in a natural way.

The Wigner description is based on the ``pseudo-distribution function", defined as
\begin{equation}
F (\bm{r},\bm{v},t) = \left(\frac{m}{2 \pi \hbar}\right)^{3} \int \rho  ( \bm{r} - \bm{\lambda} / 2, \bm{r} + \bm{\lambda} / 2,t) \exp \left[\frac{ i m \bm{v} \cdot \bm{\lambda}}{\hbar }\right] d\bm{\lambda},
\label{wignerfunction}
\end{equation}
where, for particles with spin 1/2, $F$ is a $2\times 2$ matrix and  $\rho $ is the density matrix of the system.  The matrix components of the density matrix $\rho ^{\eta \eta '} (\bm{r},\bm{r}',t) $ where $\eta =  \uparrow , \downarrow  $, are given by
\begin{equation}
\rho ^{\eta \eta '} (\bm{r},\bm{r}') = \sum _{\mu} \Psi_{\mu}^{\eta}(\bm{r},t) \Psi_{\mu}^{\eta ' *}(\bm{r}',t).
\label{matrix density}
\end{equation}
In order to study the macroscopic properties of the system, it is convenient to project $F$ onto the Pauli basis set \cite{barletti_03,morandi_JPA_11}
\begin{equation}
 F= \frac{1}{2}\sigma _0 f_0 + \frac{1}{\hbar}\bm{f} \cdot \bm{\sigma},
\label{change basis wigner function}
\end{equation}
where
\begin{equation}
f _{0} = \textrm{tr} \left\{ F \right\}  = f ^{\uparrow \uparrow} + f ^{\downarrow \downarrow}, ~~~~
\bm{f}   = \frac{\hbar}{2} \textrm{tr} \left( F \bm \sigma \right)
\label{change basis wigner function}
\end{equation}
and tr denotes the trace.
With this definition, the particle density $n $ and the spin polarization $\bm{S} $ of the electron gas are easily expressed by the moments of the pseudo-distribution functions  $f_0$ and $\bm f$:
\begin{eqnarray}
n(\bm{r},t)
&=&\sum _{\mu}  \left|  \Psi_{\mu}^\dagger (\bm{r},t) \right| ^2
=
\int f_{0} (\bm{r},\bm{v}, t) d\bm{v}, \label{def n} \\
\bm S (\bm{r},t)
&=&
\frac{\hbar}{2} \sum_{\mu}  \Psi_{\mu}^\dagger(\bm{r},t)\bm \sigma  \Psi_{\mu}(\bm{r},t)
=
\int \bm f (\bm{r},\bm{v}, t)  d\bm{v}.\label{def S}
\end{eqnarray}

In this representation, the Wigner functions have a clear physical interpretation: $f_{0}$ is related to the total electron density (in phase space), whereas $f_{i}$ ($i=x,y,z$) is related to the spin polarization in the direction $i$.
In other words, $f_{0}$ represents the probability to find an electron at one point of the phase space at a given time, while $f_{i}$ represents the probability to have a spin-polarization probability in the direction $i$ for this electron.
Using Eq. \eqref{ks equation}, some straightforward calculations lead to the quantum evolution equations for the Wigner functions
\begin{eqnarray}
&&\frac{\partial f_{0}}{\partial t} + \bm{v} \cdot \bm{\nabla}_{\bm{r}}f_{0}
+\mathcal{Q}_V [f_0]+\mu_{B} \mathcal{Q}_{B_i} [f_i] = 0, \label{f0_evo}\\
&& \nonumber \\
&&\frac{\partial f_{i}}{\partial t} + \bm{v} \cdot \bm{\nabla}_{\bm{r}}f_{i}
+\mathcal{Q}_V [f_i] +\mu_{B} \mathcal{Q}_{B_i} [f_0]+  \mu_{B} \epsilon_{ijk} \mathcal{Q}_{B_j} [f_k]=0.
\label{falpha_evo}
\end{eqnarray}
Here, $\epsilon_{irl}$ is the Levi-Civita symbol, and we used the Einstein summation convention on repeated indices. Further, we defined the pseudo-differential operator
\begin{align}
\mathcal{Q}_R [f]=&  \left(\frac{m }{ 2 \pi \hbar }\right)^{3} \int  \frac{R( \bm{r} +\bm{\lambda} / 2,t) - R( \bm{r} - \bm{\lambda} / 2,t)}{i \hbar}  \times \nonumber \\ & \hspace{3cm} f (\bm{r},\bm{v}',t) \exp \left[\frac{ i m \left(\bm{v} - \bm{v}'\right) \cdot \bm{\lambda}}{\hbar }\right]\;  d\bm{\lambda} ~ d\bm{v}',
\end{align}
where $R$ can be either the scalar potential $V$ or one of the components of the magnetic field $B_i$.
Equations \eqref{f0_evo}--\eqref{falpha_evo} describe the particle motion in a fully quantum-mechanical fashion. The integral form of the operator $\mathcal{Q}$, which generalizes the classical force operator, makes the study of such a system particularly challenging \cite{Querlioz_13,morandi_PhysRevB_09,mor_JPA_10,morandi_CAIM_10}.

In order to obtain a semiclassical approximation, we take the classical limit of Eqs. \eqref{f0_evo}--\eqref{falpha_evo} and only keep the first the correction to the Vlasov motion induced by the Zeeman-like interaction between the spin and the magnetic field.
A simple approach to derive the classical limit is to expand the operator $\mathcal{Q}$ in a power series of $\hbar$. At zeroth order, the equations for $f_0$ and $f_i$ decouple, so that one can study the particle motion irrespective from the spin degrees of freedom, and the equation for $f_0$ becomes identical to the classical Vlasov equation.
Up to first order in $\hbar$, we obtain
\begin{align}
&\frac{\partial f_{0}}{\partial t} + \bm{v} \cdot \bm{\nabla}_{\bm{r}}f_{0}
- \frac{e}{ m } \left(\bm{E} + \bm{v} \times \bm{B} \right) \cdot \bm{\nabla}_{\bm{v}} f_{0} - \frac{e }{m^{2}} \sum_{i} \bm{\nabla}_{\bm{r}} B_{i} \cdot \bm{\nabla}_{\bm{v}} f_{i}  =  0,  \label{f0_evo_vlasov}  \\
&\frac{\partial f_{i}}{\partial t} + \bm{v} \cdot \bm{\nabla}_{\bm{r}}f_{i}
- \frac{e}{ m } \left[ \left(\bm{E} + \bm{v} \times \bm{B} \right)\cdot \bm{\nabla}_{\bm{v}} f_{i} - \left( \bm{f} \times \bm{B} \right) _{i} \right] - \frac{\mu_{B} \hbar}{2m}  \bm{\nabla}_{\bm{r}} B_{i} \cdot \bm{\nabla}_{\bm{v}} f_{0}
  =0, \label{falpha_evo_vlasov}
\end{align}
where the electric field $\bm{E}$ is given by $\bm{\nabla}V =  e\bm{E}$.

We note that the $\hbar\rightarrow 0$ limit of the quantum system \eqref{f0_evo}--\eqref{falpha_evo} does not yield the Lorentz force $\bm{v} \times \bm{B}$. This is because in the Schr\"odinger-Pauli equation \eqref{ks equation} we defined, for simplicity, the kinetic energy as $\widehat{\mathbf{p}}^2/2m$, instead of the correct expression $(\widehat{\mathbf{p}}+e\bm{A})^2/2m$, where $\bm{A}$ is the vector potential such that $\bm B=\nabla \times \bm{A}$. (This is an often-used approximation in condensed matter physics, which amounts to neglecting the effect of the magnetic field on the orbital motion). Using the correct expression
[and replacing $\bm v$ with $\bm p$ in Eq. \eqref{wignerfunction}]
leads to considerably more complicated forms for the Wigner evolution equations \eqref{f0_evo}--\eqref{falpha_evo}. Nevertheless, it can be proven \cite{arnold} that in the limit $\hbar\rightarrow 0$, one does obtain the Vlasov equations \eqref{f0_evo_vlasov}--\eqref{falpha_evo_vlasov}.

Equations \eqref{f0_evo_vlasov}--\eqref{falpha_evo_vlasov} constitute the Vlasov model that we will use throughout the rest of this paper. Compared to a particle without spin, the evolution is described by a $2 \times 2$ matrix of phase-space functions. This reflects the quantum nature of the spin variable, which is a two-component vector in a Hilbert space. In contrast, the orbital degrees of freedom are treated in a completely classical way.

According to Eq. \eqref{def n}, the scalar distribution $f_0$ provides the particle density, whereas the vector distribution $\bm{f}$ yields the spin polarization as defined in Eq. \eqref{def S}. One can prove the following bound:
\begin{equation}
\left| \bm S (\bm r,t) \right| \leq n(\bm r,t)\frac{\hbar}{2}.  \label{S leq n}
\end{equation}
Equation \eqref{S leq n} is a direct consequence of the following property of the density matrix: $\textrm{tr} \left( \rho^{2} \right) \leq 1$. The equality holds true for a pure state or for a fluid where all the spins are aligned along the same direction (fully spin-polarized state).

The term $ \bm{f} \times \bm{B} $ in Eq. \eqref{falpha_evo_vlasov} represents the spin precession operator (rotation of the spin phase-space density $\bm{f}$ around the magnetic field). The remaining terms couple the equations for $f_0$ and $\bm{f}$. Such coupling exists only in the presence of an inhomogeneous magnetic field ($\nabla_\mathbf{r} B_i \neq 0$) and is a truly quantum effect. These terms reflect the force exerted on a magnetic dipole by an inhomogeneous magnetic field, which is at the basis of Stern-Gerlach-type experiments.

The Vlasov equations \eqref{f0_evo_vlasov}--\eqref{falpha_evo_vlasov} should also be compared to the kinetic model proposed by Zamanian et al. \cite{zamanian_NJP10}, where the spin is introduced as a classical {\em independent} variable on a par with the position and the velocity of a particle. Thus, the distribution function evolves in an extended phase space $(\bm{r},\bm{v},\bm{s}$). This is in contrast with our approach, where the spin is treated as a fully quantum variable (evolving in a two-dimensional Hilbert space).
Nevertheless, it can be proven that the two sets of equations are equivalent. This can be done by integrating the equations of Ref. \cite{zamanian_NJP10} in the spin variable $\bm{s}$  \footnote{{\new Such an equivalence may seem surprising, as by integrating in the spin variable
some information should invariably be lost. However, the distribution functions used by Zamanian et al. constitute only a subset of all possible functions in the extended phase space, as is apparent from Eq. (27) in Ref. \cite{zamanian_NJP10}. Within this subset, our ($2\times 2$ matrix) $f(\bm{r},\bm{v})$ and their (scalar) $f_Z(\bm{r},\bm{v},\bm{s})$ contain the same information and the two models are indeed equivalent.}}, and using the correspondence relations between our distribution functions $f_0(\bm{r},\bm{v},t)$ and $f_i(\bm{r},\bm{v},t)$ and the scalar distribution used by Zamanian et al. \cite{zamanian_NJP10} $f_Z(\bm{r},\bm{v},\bm{s},t)$, namely:
\[
f_0 = \int f_Z d^2 {\bm s}~,~~~ f_i = 3\int s_i f_Z d^2 {\bm s}.
\]

\section{Hydrodynamic model with spin}\label{sec:hydro}

In this Section, starting from Eqs. \eqref{f0_evo_vlasov}--\eqref{falpha_evo_vlasov}, we derive the hydrodynamic evolution equations by taking velocity moments of the phase-space distribution functions.
In addition to the particle density and spin polarization [Eqs.
\eqref{def n} and \eqref{def S}], we define the following macroscopic quantities
\begin{eqnarray}
\bm{u} &=& \frac{1}{n}\int \bm{v} f_{0}d\bm{v},\label{def u} \\
J^{S}_{i\alpha}&=& \int v_{i} f_{\alpha}d\bm{v},\label{def J}\\
P_{ij}&=& m\int w_{i} w_{j} f_{0}d\bm{v},\label{def P}\\
\Pi_{ij\alpha} &=& m  \int v_{i} v_{j}  f_{\alpha}  d\bm{v},\label{def Pi}\\
Q_{ijk} &=& m \int w_{i} w_{j} w_{k} f_{0} d\bm{v},\label{def Q}
\end{eqnarray}
where we separated the mean fluid velocity $\bm u$ from the velocity fluctuations $\bm{w} \equiv \bm{v} - \bm{u}$.
Here, $P_{ij}$ and $Q_{ijk} $ are respectively the pressure and the generalized energy flux tensors. They coincide with the analogous definitions for spinless fluids with probability distribution function $f_0$.
The spin-velocity tensor $J_{i \alpha}^S$ represents  the mean fluid velocity along the $i-$th direction of the $\alpha-$th spin polarization vector, while $\Pi_{ij\alpha}$ represents the corresponding spin-pressure tensor
\footnote{Strictly speaking a pressure tensor should be defined in terms of the velocity fluctuations $w_i w_j$, but this would unduly complicate the notation. Thus, we stick to the above definition of $\Pi_{ij\alpha}$ while still using the term ``pressure" for this quantity.}.

The evolution equations for the above fluid quantities are easily obtained by the straightforward integration of Eqs. \eqref{f0_evo_vlasov}-\eqref{falpha_evo_vlasov} with respect to the velocity variable. We obtain (here and in the following, we again use Einstein's summation convention):
\begin{align}
&\frac{\partial n}{\partial t} + \bm{\nabla}_{\bm{r}} \cdot \left(n\bm{u}\right)   = 0,  \label{f0_continuity} \\
&\frac{\partial S_{\alpha}}{\partial t} + \partial_{i} J^{S}_{i \alpha} + \frac{e}{ m } \left( \bm{S} \times \bm{B} \right) _{\alpha} =  0, \\
\label{falpha_continuity}
&\frac{\partial u_{i} }{\partial t} + u_{j} (\nabla_{j} u_{i}) + \frac{1}{nm} \nabla_{j} P_{ij} + \frac{e}{m} \left[E_{i} + \left(\bm{u} \times \bm{B} \right)_{i}  \right]+\frac{e}{nm^{2}}  S_{\alpha} \left(\partial_{i} B_{\alpha} \right)
= 0,  \\
&\frac{\partial J^{S}_{i \alpha}}{\partial t} + \partial_{j} \Pi_{ij\alpha} + \frac{e E_{i}}{ m }S_{\alpha} + \frac{e}{ m } \epsilon_{jki} B_{k} J^{S}_{j\alpha} + \frac{e}{ m } \epsilon_{jk\alpha} B_{k} J^{S}_{ij} + \frac{\mu_{B} \hbar}{2m}  \left(\partial_{i} B_{\alpha} \right) n =  0,\label{evol Js}\\
&\frac{\partial P_{ij}}{\partial t}  + u_{k}  \partial_{k} P_{ij}  + P_{jk} \partial_{k} u_{i} +  P_{ik} \partial_{k} u_{j}  +  P_{ij} \partial_{k}  u_{k}    + \partial_{k} Q_{ijk} + \frac{e}{m} \big{[} \epsilon_{lki} B_{k} P_{jl}  \nonumber \\
&\hspace{1cm} + \epsilon_{lkj} B_{k} P_{il} \big{]}   + \frac{e}{m^{2}} \sum_{\alpha} \left[   \partial_{i} B_{\alpha}  \left( J^{S}_{j \alpha} - S_{\alpha} u_{j} \right) +   \partial_{j} B_{\alpha}  \left(J^{S}_{i \alpha} - S_{\alpha} u_{i} \right) \right] = 0,
\label{eq_presure}
\end{align}

Other sets of hydrodynamic equations for spin-1/2 particles were derived by Brodin and Marklund \cite{Brodin} using a Madelung transformation on the Pauli wave function. The resulting model is much more cumbersome than the above system \eqref{f0_continuity}-\eqref{eq_presure}, and it is hard to identify the physical meaning of each term in their equations.
A different hydrodynamic theory was derived by Zamanian et al. \cite{zamanian_POP10} from a Vlasov equation that includes the spin as an independent variable \cite{zamanian_NJP10}. Their equations are very similar to ours. The main difference is that, in the equations of Ref. \cite{zamanian_POP10}, each quantity (including the spin polarization) is transported by a fluid element traveling with the mean fluid velocity $\mathbf{u}$. In other words, the convective derivative is always $D_t = \partial_t + \mathbf{u} \cdot \nabla$. In contrast, in our equations \eqref{f0_continuity}-\eqref{eq_presure}, only the spinless quantities (velocity, pressure) are transported by the fluid velocity, whereas the spin quantities ($S_\alpha$, $J^S_{i \alpha}$) are not.
{\new
However, it can be shown that our fluid equations \eqref{f0_continuity}-\eqref{eq_presure} are equivalent to those of Ref. \cite{zamanian_POP10}. The apparent discrepancy in the two sets of fluid equations arises mainly from the different definitions of the velocity moments in the two approaches.
}

As is always the case for hydrodynamic models, some further hypothesis is needed to close the above set of equations \eqref{f0_continuity}-\eqref{eq_presure}. In the next Section, we will deal with the closure problem by resorting to a maximum entropy principle (MEP) -- an approach that has been developed for spinless systems and that can be straightforwardly generalized to our case of a fluid with spin.

In order to fix the ideas before addressing the general framework of the MEP, we discuss an intuitive closure relation that arises naturally from the equations. In Sec. \ref{3_moment_section}, this intuitive approach will be justified rigourously on the basis of the MEP, and then overcome in Sec. \ref{4_moment_section}.
We first note that, by definition, the following equation is always satisfied:
$\int w_{i} f_{0} d\bm{v} = 0$.
The same is not true, however, for the expression obtained by replacing $f_0$ with $f_{\alpha}$ in the preceding integral. If we {\em assume} that such a quantity indeed vanishes, i.e.
$\int w_{i} f_{\alpha} d\bm{v} =0$,
we immediately obtain that
\begin{align}
J^{S}_{i\alpha} = u_{i} S_{\alpha}. \label{int clos}
\end{align}
The physical interpretation of the above equation is that the spin of a particle is simply transported along the mean fluid velocity.
This is of course an approximation that amounts to neglecting some spin-velocity correlations \cite{zamanian_POP10}.

With this assumption,  Eq. \eqref{evol Js} and the definition of the spin-pressure $\Pi_{ij\alpha}$ are no longer necessary. The system of fluid equations simplifies to
\begin{align}
& \frac{\partial n}{\partial t} + \bm{\nabla}_{\bm{r}} \cdot \left(\bm{u} n\right)    =     0, \label{density_first_closure}\\
&\frac{\partial S_{\alpha}}{\partial t} + \partial_{i} \left(u_{i} S_{\alpha}\right) + \frac{e}{ m } \left( \bm{S} \times \bm{B} \right) _{\alpha}   = 0, \label{spin_first_closure} \\
&\frac{\partial u_{i} }{\partial t} + u_{j} (\nabla_{j} u_{i}) + \frac{1}{nm} \nabla_{j} P_{ij} + \frac{e}{m} \left[E_{i} + \left(\bm{u} \times \bm{B} \right)_{i}  \right]+\frac{e}{nm^{2}}  S_{\alpha} \left(\partial_{i} B_{\alpha} \right)   =  0, \label{velocity_first_closure} \\
&\frac{\partial P_{ij}}{\partial t}  + u_{k}  \partial_{k} P_{ij}  + P_{jk} \partial_{k} u_{i} +  P_{ik} \partial_{k} u_{j}  +  P_{ij} \partial_{k}  u_{k}    + \partial_{k} Q_{ijk} \nonumber \\
&\hspace{2cm} + \frac{e}{m} \big{[} \epsilon_{lki} B_{k} P_{jl}   + \epsilon_{lkj} B_{k} P_{il} \big{]}  = 0,
\label{eq_presure2}
\end{align}
Interestingly, in Eq. \eqref{spin_first_closure} the spin polarization is now transported by the fluid velocity $\mathbf{u}$, as in the model of Zamanian et al. \cite{zamanian_POP10}.

We note that in Eqs. \eqref{density_first_closure}--\eqref{eq_presure2} we have already closed [thanks to Eq. \eqref{int clos}] the spin-dependent part of the equations.
In order to complete the closure procedure, one can proceed in the same way as is usually done for spinless fluids, for instance by supposing that the system is isotropic and adiabatic.
The isotropy condition imposes that $P_{ij} = (P/3) \delta_{ij}$ where $\delta_{ij}$ is the Kronecker delta, while the adiabatic condition requires that the heat flux $Q^{th}_{i} = m \int \bm{w}^{2} w_{i} f_{0} d\bm{v}$ vanish. In this case, one can prove that the pressure takes the usual form for the equation of state of an adiabatic system, i.e., $P ={\rm const.} \times n^{\frac{D+2}{D}}$ ($D$ is the dimensionality of the system), which replaces Eq. \eqref{eq_presure2}.
In summary, Eqs. \eqref{density_first_closure}-\eqref{velocity_first_closure}, together with the preceding expression for the pressure, constitute a closed system of hydrodynamic equations with spin.

\section{Fluid closure: Maximum entropy principle}\label{sec:MEP}

The maximum entropy principle is a well-developed theory that has been successfully applied to various areas of gas, fluid, and solid-state physics \cite{Ali_12,Trovato_10,Romano_01,Anile_95}.
The underlying assumption of the MEP is that, at equilibrium, the probability distribution function is given by the most probable microscopic distribution  (i.e., the one that maximizes the entropy) compatible with some macroscopic constraints. The constraints are generally given by the various velocity moments, i.e., the local density, mean velocity, and temperature.
From a mathematical point of view, this procedure leads to a constrained maximization problem.

In order to illustrate the application of the MEP theory to a spin system, we write the Hamiltonian in a more general way
\begin{equation}
  \mathcal{H}= h_0 (\bm{r},\bm{v})\sigma_0 +\bm{h}(\bm{r},\bm{v}) \cdot \bm{\sigma},
\end{equation}
where $h_0 $ and $\bm{h} $ are
functions of the particle position $\bm{r}$ and velocity $\bm{v} \equiv (\bm{p}+e\bm{A})/m$.
In our case
\begin{align}
  h_0 =&   m \frac{|\bm{v}|^2}{2}   + V    ,\\
 \bm{h} =& \mu _{B}  \bm{B}  .
\end{align}
In order to simplify the notation, we denote the fluid moments by
\begin{equation}
  \bm{m}_i(\bm{r}) = \textrm{tr} \int  \bm \chi_{i} F  d\bm{v}, \label{mom cont}
\end{equation}
where $\chi _{i} $ is the function associated with the $i-$th moment. Thus, the definitions \eqref{def n}--\eqref{def S} and \eqref{def u}--\eqref{def Q} correspond to
\begin{equation}
\bm{m}= \left(
   \begin{array}{c}
     n \\
     \bm{S} \\
     \bm{u} \\
   J^{S}_{i \alpha} \\
    \vdots \\
   \end{array}
 \right) ; ~~~~~~ \bm \chi =  \left(
   \begin{array}{c}
     1 \\
     \bm{\sigma} \\
     \bm{v} \\
      v_i \sigma _\alpha \\
     \vdots \\
   \end{array}
 \right).
\end{equation}
The relevant entropy  density is
\begin{equation}
 s(F)=\left\{
 \begin{array}{ll}
               k_B ~\textrm{tr} \left\{ F \log F -F \right\} &  ~~\textrm{(M--B)}\\[2mm]
               k_B~ \textrm{tr} \left\{ F \log F+ (1-F) \log (1-F) \right\} & ~~\textrm{(F--D),}
              \end{array}\right.
\end{equation}
where we distinguished between Maxwell-Boltzmann (M--B) and Fermi-Dirac (F--D) statistics. The MEP assumes that the phase-space distribution function $F$ is the extremum of the free-energy functional
\begin{eqnarray}
  \mathcal{E}&=&
  \textrm{tr}  \int \left[ T s(F)+  \mathcal{H}'   F \right] d\bm{v} d\bm{r} - \int \lambda _i (\bm{r}) m_i(\bm{r}) d\bm{r},
\end{eqnarray}
where we defined $ \mathcal{H}' =  \mathcal{H} +\lambda _i (\bm{r})  \bm \chi_i$, $T$ is the temperature and the functions $\lambda_i$ are the Lagrange multipliers. The $\lambda_i$ constitute a set of independent functions that are used to parameterize the equilibrium distribution $F^{eq}$. A major technical difficulty of the MEP method is to express the $\lambda_i$ set  in terms of $\bm{m}$ in a closed form. This point will be illustrated in details in the following paragraphs. The total variation (Lie derivative) of $\mathcal{E}$ gives
\begin{eqnarray}
 \delta \mathcal{E}&=& \delta \lambda_i \frac{\delta}{\delta \lambda_i} \mathcal{E}  + \delta F\frac{\delta}{\delta F} \mathcal{E}. \label{lie der}
\end{eqnarray}
The local equilibrium distribution $F^{eq}$ corresponds to the extremum
$\delta \mathcal{E}(F^{eq})= 0$.
It is easy to verify that the variation with respect the Lagrange multipliers [the first term of the right hand side of Eq. \eqref{lie der}] gives Eq.  \eqref{mom cont}.

The equilibrium distribution is formally obtained by taking the variation of $\mathcal{E}$ with respect to $F$
\begin{equation}
   \delta F\frac{\delta\mathcal{E}}{\delta F}  = \textrm{tr} \int \left[T \frac{\delta s}{\delta F}+  \mathcal{H}' \right]\delta F  d\bm{v} d\bm{r}.
\end{equation}
Setting $\delta\mathcal{E}/\delta F=0$, yields
\begin{equation}
F^{eq} =\left\{ \begin{array}{ll}
a \exp\left( - \beta \mathcal{H}' \right) & ~~ \textrm{(M--B)}\\
a \left[\exp\left(\beta \mathcal{H}' \right) + 1\right]^{-1} &~~\textrm{(F--D),}
\end{array}\right.
\label{statistic}
\end{equation}
where $a$ is a constant and $\beta=1/(k_B T)$. Equation \eqref{statistic} is a very general result that holds irrespectively of the number and the type of moments that are being considered. For every specific choice of the moments to be preserved, the explicit form of the local equilibrium function $F^{eq}$ can be constructed from Eq. \eqref{statistic}. In order to illustrate the results for a fluid with spin, in the next sections we shall consider various models characterized by a different number of fluid moments (three or four) and by the use of the M--B or F--D statistics.

\section{Three-moment closure} \label{3_moment_section}
To begin with, we consider a simplified situation where only three fluid moments (density $n$, mean velocity $\bm{u}$, and spin polarization $ \bm{S}$) are kept, that is:
\begin{equation}
\bm{m}= \left(
   \begin{array}{c}
     n \\
     \bm{S} \\
     \bm{u} \\
   \end{array}
 \right).
\end{equation}
It is convenient to write the hamiltonian $\mathcal{H'}$ in the following way
\begin{equation}
\mathcal{H'} = h_{0}' + \bm{h'}\cdot \bm{\sigma} = \frac{m}{2} \left( \bm{v } - \bm{v_{0}} \right)^{2} + \lambda_{0} + \bm{\lambda_{S}} \cdot \bm{\sigma},
\end{equation}
where the Lagrange multipliers $\lambda_{0}$, $\bm{\lambda_{S}}$ and $\bm{v_{0}}$ (seven scalar quantities in total) are associated respectively to the density, the spin polarization vector, and the mean velocity.
We then evaluate the equilibrium distribution for the M--B and F--D statistics.

\subsection{Maxwell-Boltzmann statistics}

We fix the normalization constant $a_0=\left(\frac{m}{2\pi \hbar} \right)^{3}$. Equation  \eqref{statistic} (for M--B statistics) gives
\begin{eqnarray}
F^{eq} &=& a_0 ~\sigma _0  e^{  - \beta h_0'}\exp \left(   -   \beta  \bm {h}'    \cdot \bm{\sigma} \right)\nonumber \\
&=&   a_0  \left[\; \sigma _0   \cosh \left( -  \beta |\mathbf{h}'|  \right)   +   \frac{\bm{h}'\cdot \bm{\sigma} }{|\bm{h}'|} \sinh \left( -  \beta |\bm{h}'|  \right)  \right]e^{  - \beta h_0'}.\label{Feq MB}
\end{eqnarray}
By calculating the moments of $F^{eq}$, we can express the fluid moments in terms of the Lagrangian multipliers. We find
\begin{eqnarray*}
n &=&  2 a_0 \Gamma(T) \exp \left(-\beta \lambda_{0} \right) \cosh \left( - \beta |\bm{\lambda_{S}}| \right), \\
\bm{S} &=&   \hbar ~a_0 ~\frac{ \bm{\lambda_{S}}}{|\bm{\lambda_{S}}|}~ \Gamma(T) \exp \left(-\beta \lambda_{0} \right) \sinh \left( - \beta |\bm{\lambda_{S}}| \right), \\
\bm{u} &=& \bm{v_{0}} ,
\end{eqnarray*}
where $\Gamma(T) =  \left(2\pi  k_{B} T /m \right) ^{3/2} $. The previous equations can be inverted:
\begin{align}
\exp \left(-\beta \lambda_{0} \right) =& \displaystyle a_0 \frac{1}{2 \Gamma(T)} \sqrt{\left( n^{2} - \frac{4 |\bm{S}|^{2}}{\hbar ^{2}}\right)},   \\
\bm \lambda_{S}   =& \displaystyle \frac{\bm S}{|\bm{S}|} \frac{k_{B}T}{2}   \ln\left( \frac{n - \frac{2|\bm{S}|}{\hbar}}{n + \frac{2|\bm{S}|}{\hbar}} \right) .\label{lagrange_multipliers_maxwell_3}
\end{align}
Note that the quantities on the right-hand side of the above expressions are real, thanks to Eq. \eqref{S leq n}.

Finally, the equilibrium distribution can be expressed in terms of the fluid moments in a simple form
\begin{eqnarray}
F^{eq} = \left( \sigma _0  n  +  \bm{\sigma} \cdot   \bm{S}    \right)\frac{1 }{\Gamma(T)}  \exp \left( - \beta \frac{m\left( \bm{v} - \bm{u}\right)^{2}}{2}\right).
\end{eqnarray}
The pressure and the spin current at equilibrium are thus given by
\begin{align}
P_{ij} &= m\;  \textrm{tr} \left( \int v_{i} v_{j} F^{ eq } d\bm{v}\right) - m n u^{2}  =   n k_{B} T \delta_{ij} \label{pressure_maxwell_3}\\
J^{S}_{i \alpha} &= S_{\alpha} u_{i}. \label{spin_current_maxwell_3}
\end{align}
Thus, considering three fluid moments and M--B statistics, leads to the standard expression for the isotropic pressure of an ideal gas, together with
the ``intuitive" closure condition \eqref{int clos} for the spin current tensor.

\subsection{Fermi-Dirac statistics}
We now consider the F--D case. After some tedious but straightforward calculations (details can be found in Appendix \ref{app_FD_3}), Eq. \eqref{statistic} gives
\begin{equation}
F^{eq}
=
\frac{a_{0}}{2}\frac{  \left( \cosh \left(\beta |\bm{h'}|\right) + \exp ^{-\beta h_{0}' }  \right) \sigma_{0} - \sinh \left(\beta h_{0}'\right) \frac{\bm{h'} \cdot \bm{\sigma}}{|\bm{h'}|} }{ \left[ \cosh \left(\beta h_{0}'\right) + \cosh \left(\beta |\bm{h'}|\right)\right]} .
\label{f_eq_FD}
\end{equation}
In the case of the F--D statistics, it is no longer possible to obtain a closed expression of $F^{eq}$ when $T > 0$.  However, for many applications of the hydrodynamic model, the assumption that the particle have zero temperature is not too restrictive. Indeed, for solid-state metallic densities, the Fermi temperature is of the order $T_F \approx 5\times 10^4 ~\rm K$, so that in the vast majority of conceivable situations $T \ll T_F$, and the zero-temperature approximation is sufficiently accurate.

We have evaluated the macroscopic moment of $F^{eq} $ in the case $T=0$. We obtain (details of the calculations are given in Appendix \ref{app_FD_3}):
\begin{eqnarray}
 n
 &=&
 \frac{4 \pi}{3}a_{0}\left( \left[ \frac{2}{m} \left( |\bm{\lambda_{S}}| + |\lambda_{0}| \right) \right]^{3/2} + \left[ \frac{2}{m} \left( |\lambda_{0}| - |\bm{\lambda_{S}}|  \right) \right]^{3/2} \right), \label{density FD} \\
 \bm{S}
 &=&
  -\frac{\hbar}{2} a_{0} \frac{\bm{\lambda ^{S}}}{|\bm{\lambda^{s}}|} \frac{4 \pi}{3}\left( \left[ \frac{2}{m} \left( |\bm{\lambda^{S}}| + |\lambda_{0}| \right) \right]^{3/2} - \left[ \frac{2}{m} \left( |\lambda_{0}| - |\bm{\lambda^{S}}|  \right) \right]^{3/2} \right),  \label{spin FD} \\
\bm{u}
&=&
\bm{v_{0}}. \label{u FD}
\end{eqnarray}
Note that, in the above expressions, the quantities under square root are nonnegative for all physically admissible states, as is shown in Appendix \ref{app_FD_3}.

As in the case of M--B statistics, we find that $J^{S}_{i\alpha} = u_{i} S_{\alpha}$. For the pressure, we obtain
\begin{eqnarray}
P
&=&
\frac{\hbar^{2}}{5m}\frac{\left(6\pi ^{2} \right)^{2/3}}{2^{5/3}} \left[\left( n - \frac{2}{\hbar}|\bm{S}|\right)^{5/3} + \left( n + \frac{2}{\hbar}|\bm{S}|\right)^{5/3} \right].\label{P FD}
\end{eqnarray}
When the spin polarization vanishes, Eq. \eqref{P FD} reduces to the usual expression of the zero-temperature pressure of a spinless Fermi gas: $P=\frac{\hbar^{2}}{5m} \left(3\pi ^{2} \right)^{2/3}  n^{5/3}$.
The modification of the spin pressure induced by the spin has a simple physical interpretation. Equation \eqref{P FD} can be interpreted as the total pressure of a plasma composed by two populations, the spin-up and the spin-down particles. Due to the Zeeman splitting, the density of the particles whose spin is parallel to the magnetic field is lower than the energy of the particles whose spin is antiparallel. Equation \eqref{P FD} shows that the two populations provide a separate contribution to the total fluid pressure.

\section{Four-moment closure} \label{4_moment_section}

As a final example, we consider the complete four-moment model:
\begin{equation}
\bm m =
\begin{pmatrix}
n \\ \bm{S} \\ \bm{u} \\ J^{S}_{i \alpha}
\end{pmatrix} ~~~~\textrm{and} ~~~~
\chi = \begin{pmatrix}
\lambda_0 \\ \bm{\lambda_S} \\ \bm{v_0}\\ \lambda^{J}_{i \alpha}
\end{pmatrix}.
\end{equation}
In this case, the hamiltonian $\mathcal H'$ becomes
\begin{equation}
\mathcal{H}' = \frac{m \left( \bm{v} - \bm{v_{0}} \right)^{2}}{2} + \lambda^{0} + \left(  \lambda^{S}_{\alpha} + \lambda^{J}_{i \alpha} v_{i}\right) \sigma_{\alpha}. \label{Hp 4m}
\end{equation}

Here, we consider a particular situation where the evaluation of the closure expressions can be obtained analytically, namely the collinear case with Maxwell-Boltzmann statistics. With the term ``collinear" we denote a fluid whose spin polarization is parallel to a fixed direction (here, the $z$ direction). In the collinear case, the Hamiltonian reduces to ${\cal H}_{\rm col}=\frac{m}{2} v^2+ \mu_B B_z \sigma_z$.
The equilibrium distribution $F^{eq}$ is given by Eq. \eqref{Feq MB} with
\begin{align}
  h_{0}' =& m\left(\bm{v} - \bm{v_{0}} \right)^{2}/2 + \lambda_{0} \label{h0_mb 4} \\
  h'_{z} =&  \lambda^{S}_{z} + \lambda^{J}_{xz} v_{x} + \lambda^{J}_{yz} v_{y}  + \lambda^{J}_{zz} v_{z}  \label{hz_mb 4}\\
  h'_{x} =&h'_{y} =0.
\end{align}
Proceeding as before, we obtain the relations between the moments and the Lagrange multipliers. The details of the calculations are given in Appendix \ref{app_MB_4}. We obtain
\begin{align}
\bm{\gamma}  =&  \frac{2n \hbar m}{\hbar ^{2} n^{2} + 4 S_{z}^{2}} \left( S_{z} \bm u  - \bm J^{S} \right), \label{begin_mb_4}\\
\bm {v_{0}}  =&   \frac{1}{\hbar ^{2} n^{2} + 4 S_{z}^{2}} \left( \hbar ^{2} n^{2} \bm u  + 4 S_{z} \bm J^{S}  \right),  \\
e^{ - \beta \lambda_{0} } =& \frac{e^{ \beta \gamma^{2} / 2m}}{\Gamma(T)} \sqrt{\left( \frac{n}{2} \right)^{2} - \left( \frac{S_{z}}{\hbar} \right)^{2}},  \\
\lambda^{S}_{z} =& \displaystyle \frac{k_{B}T}{2}   \ln\left( \frac{n - \frac{2|\bm{S}|}{\hbar}}{n + \frac{2|\bm{S}|}{\hbar}} \right)  - \bm{\gamma} \cdot \bm{v_{0}}. \label{end_mb_4}
\end{align}
In order to simplify the notation, we defined $\gamma_{i} = \lambda^{J}_{iz}$ and $J^{S}_{iz} = J^{S}_{i}$.

We can now calculate the equilibrium distribution function:
\begin{align}
F^{eq}  =& \frac{ e^{\beta \bm{\gamma}^{2}/2m}}{\Gamma(T)} e^{-\beta m \left( \bm{v} - \bm{v_{0}} \right)^{2}/2} \bigg{\{} \sigma_0 \left[ n \cosh \left( \beta \bm{\gamma} \cdot \left( \bm{v} - \bm{v_{0}} \right) \right) - \frac{2 S_{z}}{\hbar} \sinh
\left( \beta \bm{\gamma} \cdot \left( \bm{v} - \bm{v_{0}} \right) \right) \right]   \nonumber \\
  &\hspace{3cm}  + \sigma_z  \left[ \frac{\hbar }{2 }n \sinh \left( -\beta \bm{\gamma} \cdot \left( \bm{v} - \bm{v_{0}} \right) \right) +    S_{z}  \cosh
\left( \beta \bm{\gamma} \cdot \left( \bm{v} - \bm{v_{0}} \right) \right) \right] \bigg{\}}.
\end{align}
Finally, we calculate the pressure tensor $P_{ij}$ and the spin pressure tensor $\Pi_{ijz}$ (details are given in the Appendix \ref{app_MB_4}). We obtain
\begin{eqnarray}
P_{ij}
&=&
  e ^{\beta \bm{\gamma}^{2}/m} \left\{ nk_{B} T\delta_{i,j}   +  mn \left(\frac{\hbar^{2} n^{2} u_{i}u_{j} + 4J^{s}_{i}J^{s}_{j}}{\hbar^{2} n^{2}  + 4 S_{z}^{2}} \right)\nonumber\right.  \\
 &~&
 \left. + 8mn S_{z} \left[ \frac{\left(J^{S}_{i}-S_{z} u_{i}\right)\left(\hbar^{2} n^{2} u_{j}+ 4 S_{z}J^{s}_{j}\right)  + \left(J^{S}_{j}-S_{z} u_{j}\right) \left(\hbar^{2} n^{2} u_{i}+ 4 S_{z}J^{s}_{i}\right)}{\left(\hbar^{2} n^{2}  + 4 S_{z}^{2}\right)^{2}}\right]\right\} \nonumber \\
&~&
-  mnu_{i}u_{j},  \label{P_maxwell_4} \\
\Pi_{ijz}
%&=&
%m\int v_{i}v_{j} f_{z}^{\textrm{eq}} d\bm{v}, \nonumber \\
&=&
  e ^{\beta \bm{\gamma}^{2}/m} \left\{ S_{z}k_{B} T \delta_{i,j}  +  mS_{z}\left(\frac{\hbar^{2} n^{2} u_{i}u_{j} + 4J^{s}_{i}J^{s}_{j}}{\hbar^{2} n^{2}  + 4 S_{z}^{2}} \right) \nonumber \label{pressure_maxwell_4} \right. \\
 &~& \left.
 + 2mn^{2} \hbar^{2} \left[ \frac{\left(J^{S}_{i}-S_{z} u_{i}\right)\left(\hbar^{2} n^{2} u_{j}+ 4 S_{z}J^{s}_{j}\right)  + \left(J^{S}_{j}-S_{z} u_{j}\right) \left(\hbar^{2} n^{2} u_{i}+ 4 S_{z}J^{s}_{i}\right)}{\left(\hbar^{2} n^{2}  + 4 S_{z}^{2}\right)^{2}}\right]\right\} .\nonumber \label{spin_pressure_maxwell_4} \\
\end{eqnarray}
It is easy to verify that Eq. \eqref{P_maxwell_4} is consistent with Eq. \eqref{spin_current_maxwell_3} in the limit $\bm{\gamma}
\to 0$.
Finally, we can write a four-moment model with collinear spin and Maxwell-Boltzmannn statistics at zero temperature:
\begin{eqnarray}
\frac{\partial n}{\partial t} &+& \bm{\nabla}_{\bm{r}} \cdot (n \bm{u}) = 0,\nonumber \\
\frac{\partial S_{z}}{\partial t} &+& \partial_{i} J^{S}_{i z} = 0, \nonumber \\
\frac{\partial u_{i} }{\partial t} &+& u_{j}\partial_{j} u_{i} + \frac{1}{nm} \partial_{j} P_{ij} + \frac{e }{ m }\left(E_{i} + \epsilon_{jki} u_{j} B_{k} \right) +\frac{e}{nm^{2}}  S_{z} \left(\partial_{i} B_{z} \right) =
0, \nonumber \\
\frac{\partial J^{S}_{i z}}{\partial t} &+& \partial_{j} \Pi_{ijz} + \frac{e E_{i}}{ m }S_{z}  + \frac{e \hbar^{2}}{4m^{2}}  \left(\partial_{i} B_{z} \right) n = 0
\end{eqnarray}
The above fluid equations, together with Eqs. \eqref{P_maxwell_4} and \eqref{spin_pressure_maxwell_4}, constitute a closed system.

\section{Conclusions}

The dynamics of a system of spin-1/2 fermions is an important issue in many areas of physics, ranging from condensed matter (electrons in bulk metals), to nanophysics (electron transport in metallic and semiconductor nanostructures) and even astrophysics (interior of white dwarfs and neutron stars).

In particular, in ultrafast spectroscopy experiments carried out on nanometric objects, the electron spin can play a crucial role, as it interacts not only with the magnetic and electric fields of the incident laser pulse, but also with the self-consistent fields generated by the electrons themselves. In view of this complex variety of possible physical mechanisms, it is necessary to develop appropriate models that take into account the spin degrees of freedom in the dynamics of the electron gas. Further, these models should not be limited to the linear response, as nonlinear effects are often important, especially for large incident laser powers.

Most existing models for the quantum electron dynamics are variations on the mean-field approximation (time-dependent Hartree equations), with various upgrades that allow one to describe electron exchange (Hartree-Fock) and correlations [density functional theory, local-density approximation (LDA)], spin effects (spin LDA), and relativistic effects (Dirac-Hartree and Dirac-Kohn-Sham equations).

The use of phase-space models is less widespread, although both the Vlasov and Wigner equations have been used in the past to study the electron dynamics in metallic nansotructures \cite{Calvayrac,metal_films_2,metal_films_3}. Some authors \cite{zamanian_NJP10, zamanian_POP10} used the Vlasov or Wigner equations in an extended phase space that includes a ``classical" spin variable.

In this paper, we derived a
a four-component Vlasov equation for a system composed of spin-1/2 fermions (typically electrons). The orbital part of the motion was assumed to be classical and therefore described by phase-space trajectories that represent the characteristics of he corresponding Vlasov equation.
In contrast, the spin degrees of freedom were treated in a completely quantum-mechanical way (two-dimensional Hilbert space). The corresponding hydrodynamic equations were derived by taking velocity moments of the phase-space distribution function. The hydrodynamic equations form an infinite hierarchy that needs to be closed on the basis of some physical hypothesis.
Here, we showed that the hydrodynamics system can be closed using a maximum entropy principle. We performed the detailed calculations for a closure with either three or four constraints on the fluid moments, for both Maxwell-Boltzmann and Fermi-Dirac statistics.

The Vlasov and fluid models that we derived in this work
should be useful, for instance, for applications to the electron dynamics in metallic nanoparticles excited with intense laser pulses, where spin and charge effects are closely intertwined.

\vskip 0.5cm
{\it \noindent Acknowledgments}\\
We thank the Agence Nationale de la Recherche, project Labex "Nanostructures in Interaction with their Environment", for financial support.

\appendix

\section{Three-moment Fermi-Dirac closure} \label{app_FD_3}

We begin by demonstrating the relation \eqref{f_eq_FD} between the equilibrium distribution $F^{eq}$ and the component of the Hamiltonian ${\cal H}' = h_{0}' \sigma_{0} + \bm{h'} \cdot \bm{\sigma'}$, where $h_{0}' = m\left(\bm{v} - \bm{v_{0}} \right)^{2}/2 + \lambda_{0}$ and $\bm{h}' = \bm{\lambda_S}$. Developing the exponential as a power series in Eq. \eqref{statistic} (F--D) and inverting the associated matrix, we obtain
\begin{eqnarray}
F^{\textrm{eq}}
&=&
a_{0}\left[ \exp\left(\beta \mathcal{H'}\right) + 1 \right]^{-1}, \nonumber \\
&=& \left(\frac{m}{2\pi\hbar}\right)^{3}  \exp\left(\beta h_{0}'\right) \left[ \cosh \left(\beta h_{0}'\right) \sigma_{0} + \cosh \left(\beta |\bm{h'}|\right) \frac{\bm{h'} \cdot \bm{\sigma}}{|\bm{h'}|}\right]^{-1} , \nonumber \\
&=&
\frac{a_{0}}{2}\frac{\left( \cosh \left(\beta |\bm{h'}|\right) + \exp ^{-\beta h_{0}' }  \right) \sigma_{0} - \sinh \left(\beta h_{0}'\right) \left(\bm{h'} \cdot \bm{\sigma}\right) / | \bm{h'}| }{ \left[ \cosh \left(\beta h_{0}'\right) + \cosh \left(\beta |\bm{h'}|\right)\right]} . \nonumber
\end{eqnarray}
In this case, we obtain the following expression for $f^{\textrm{eq}}_{0}$ and $f^{\textrm{eq}}_{i}$ :
\begin{equation*}
f^{\textrm{eq}}_{0} =a_{0}\frac{\cosh \left(\beta |\bm{h'}|\right) + \exp ^{-\beta h_{0}' } }{\cosh \left(\beta h_{0}'\right) + \cosh \left(\beta |\bm{h'}|\right)}~~\textrm{and}~~f^{\textrm{eq}}_{i} = - \frac{a_{0}~\hbar}{2} \frac{\sinh \left(\beta |\bm{h'}|\right) h'_{i} /  | \bm{h'}| }{\cosh \left(\beta h_{0}'\right) + \cosh \left(\beta |\bm{h'}|\right)}.
\end{equation*}
These expressions cannot be integrated analytically over the velocity space. To obtain a treatable model, we assume that the electron gas is at zero temperature, i.e. $\beta \rightarrow \infty$. We start by calculating the density
\begin{eqnarray}
n &=&
\lim _{\beta \rightarrow \infty} \int f_{0}^{eq} d\bm{v}
= a_{0} \lim _{\beta \rightarrow \infty} \int \frac{e^{\beta |\bm{h'}|} + 2e ^{-\beta h_{0}' }}{e^{\beta h_{0}' }+e^{-\beta h_{0}' }+e^{\beta |\bm{h'}|}} d\bm{v}
\nonumber \\
&=&
a_{0}\lim _{\beta \rightarrow \infty} \left[ \int  \frac{1}{ 1 + e^{\beta \left( h_{0}' - |\bm{h'}| \right) } + e^{-\beta \left( h_{0}' + |\bm{h'}|\right)} } d\bm{v} + 2\int  \frac{1}{ 1 + e^{2\beta h_{0}' } +e^{\beta \left( h_{0}' + |\bm{h'}|\right)} } d\bm{v} \right]. \nonumber
\end{eqnarray}
We call $n_{1}$ and $n_{2}$ respectively
the limit for $\beta \to \infty$ of
the first and the second integral in the above expression, such that $n = n_{1} + n_{2}$. One can show that
\begin{eqnarray}
n_{1} &=&  4\pi a_{0} \lim _{\beta \rightarrow \infty} \int_{0}^{+\infty} \frac{v^{2}}{1 + \exp [\beta \left( \frac{m}{2} v^{2} + \lambda_{0} - |\bm{\lambda^{S}}| \right) ] + \exp[-\beta \left( \frac{m}{2} v^{2} + \lambda_{0} - |\bm{\lambda^{S}}| \right) ] } dv \nonumber \\
&=&
\left\{
\begin{array}{lcl}
\frac{4 \pi}{3} a_{0} \left[ \frac{2}{m} \left( |\bm{\lambda^{S}}| - |\lambda_{0}| \right) \right]^{3/2} ~~~&\textrm{if}&~~~ 0 < \lambda_{0} < |\bm{\lambda^{S}}| \\
0~~~&\textrm{if}&~~~ \lambda_{0} > |\bm{\lambda^{S}}|   \\
\frac{4 \pi}{3} a_{0} \left[ \frac{2}{m} \left( |\bm{\lambda^{S}}| + |\lambda_{0}| \right) \right]^{3/2} ~~~&\textrm{if}&~~~ -|\bm{\lambda^{S}}| < \lambda_{0} < 0 \\
\frac{4 \pi}{3} a_{0}\left( \left[ \frac{2}{m} \left( |\bm{\lambda^{S}}| + |\lambda_{0}| \right) \right]^{3/2} - \left[ \frac{2}{m} \left( |\lambda_{0}| - |\bm{\lambda^{S}}|  \right) \right]^{3/2} \right) ~~~&\textrm{if}&~~~  \lambda_{0} < -|\bm{\lambda^{S}}| \\
\end{array}
\right. \nonumber \\ \nonumber \\
n_{2} &=&  8\pi a_{0}\lim _{\beta \rightarrow \infty} \int_{0}^{+\infty} \frac{v^{2}}{1 + \exp ^{2\beta \left( \frac{m}{2} v^{2} + \lambda_{0}\right) } + \exp ^{\beta \left( \frac{m}{2} v^{2} + \lambda_{0} + |\bm{\lambda^{S}}| \right) } } dv \nonumber \\
&=&
\left\{
\begin{array}{lcl}
0~~~&\textrm{if}&~~~ \lambda_{0} >- |\bm{\lambda^{S}}|  \nonumber \\
\frac{8 \pi}{3} a_{0} \left[ \frac{2}{m} \left(  |\lambda_{0}| - |\bm{\lambda^{S}}| \right) \right]^{3/2} ~~~&\textrm{if}&~~~ \lambda_{0} <- |\bm{\lambda^{S}}| \nonumber
\end{array}
\right.  \nonumber
\end{eqnarray}
For $\bm{S}$ we obtain
\begin{eqnarray}
S_{i}
&=&
 \lim _{\beta \rightarrow \infty} \int f_{i} d\bm{v}
=  -\frac{\hbar}{2} a_{0} \frac{\lambda ^{S}_{i}}{|\bm{\lambda^{S}}|}\lim _{\beta \rightarrow \infty} \int \frac{e^{\beta |\bm{h'}|} }{e ^{\beta h_{0}' }+e ^{-\beta h_{0}' }+e^{\beta |\bm{h'}|}} d\bm{v} =  -\frac{\hbar}{2} a_{0} \frac{\lambda ^{S}_{i}}{|\bm{\lambda^{S}}|} n_{1}. \nonumber
\end{eqnarray}

In the case where $\lambda_{0} >- |\bm{\lambda^{S}}|$, we have the following relation between $\bm{S}$ and $n$: $ |\bm{S}| =  \frac{\hbar}{2} n $. Comparing with Eq. \eqref{S leq n}, we notice that we are in the limit of pure states. If we consider the case where  $\lambda_{0} <- |\bm{\lambda^{S}}|$, we obtain
\begin{eqnarray}
n &=&
 \frac{4 \pi}{3} a_{0}\left( \left[ \frac{2}{m} \left( |\bm{\lambda^{S}}| + |\lambda_{0}| \right) \right]^{3/2} + \left[ \frac{2}{m} \left( |\lambda_{0}| - |\bm{\lambda^{S}}|  \right) \right]^{3/2} \right),  \nonumber \\
 \bm{S}
 &=&
  -\frac{\hbar}{2} a_{0} \frac{\bm{\lambda ^{S}}}{|\bm{\lambda^{s}}|} \frac{4 \pi}{3}\left( \left[ \frac{2}{m} \left( |\bm{\lambda^{S}}| + |\lambda_{0}| \right) \right]^{3/2} - \left[ \frac{2}{m} \left( |\lambda_{0}| - |\bm{\lambda^{S}}|  \right) \right]^{3/2} \right),  \nonumber \\
\bm{u}
&=&
\bm{v_{0}}.  \nonumber
\end{eqnarray}
It is obvious that in this case we have $|\bm{S}| \leq  \frac{\hbar}{2} n$, which is in agreement with Eq. \eqref{S leq n} and corresponds to admissible physical solutions (quantum mixed states).
We are now able to extract the following relation between the Lagrange multipliers and the fluid moments:
\begin{equation}
 |\lambda_{0}| \pm |\bm{\lambda^{S}}| = \left(\frac{2\pi\hbar}{m}\right)^{2}\frac{m}{2}\left( \frac{3}{8\pi} \right)^{2/3} \left( n \mp \frac{2}{\hbar}|\bm{S}|\right)^{2/3}. \nonumber \\
\end{equation}

The next step is to calculate the pressure $ P_{ij} = m  \int v_{i}v_{j} f^{eq}_{0} d\bm{v} - mn u_{i} u_{j}$. By using parity arguments, we deduce that the pressure must be isotropic. Thus, we obtain
\begin{eqnarray}
P
&=&
 \frac{m}{3} \int \bm{v}^{2} f^{eq}_{0} d\bm{v} - mn \bm{u}^{2} \nonumber \\
&=&
\frac{4\pi m}{3} a_{0} \bigg{[} \lim _{\beta \rightarrow \infty} \int_{0}^{+\infty} \frac{v^{4}}{1 + \exp[\beta \left( \frac{m}{2} v^{2} + \lambda_{0} - |\bm{\lambda^{S}}| \right) ] + \exp [-\beta \left( \frac{m}{2} v^{2} + \lambda_{0} - |\bm{\lambda^{S}}| \right) ] } dv \nonumber \\
&~& + 2 \lim _{\beta \rightarrow \infty} \int_{0}^{+\infty} \frac{v^{2}}{1 + \exp [2\beta \left( \frac{m}{2} v^{2} + \lambda_{0}\right) ] + \exp [\beta \left( \frac{m}{2} v^{2} + \lambda_{0} + |\bm{\lambda^{S}}| \right)] } dv \bigg{]} \nonumber \\
&=&
\frac{4\pi m}{3}\frac{a_{0} }{5} \left( \left[ \frac{2}{m} \left( |\bm{\lambda^{S}}| + |\lambda_{0}| \right) \right]^{5/2} + \left[ \frac{2}{m} \left( |\lambda_{0}| - |\bm{\lambda^{S}}|  \right) \right]^{5/2} \right) \nonumber \\
&=&
\frac{\hbar^{2}}{5m}\frac{\left(3\pi ^{2} \right)^{2/3}}{2} \left[\left( n - \frac{2}{\hbar}|\bm{S}|\right)^{5/3} + \left( n + \frac{2}{\hbar}|\bm{S}|\right)^{5/3} \right].  \nonumber
\end{eqnarray}
As to the spin current $J^{S}_{i\alpha} = \int v_{i} f_{\alpha} d\bm{v}$, we notice directly, again by parity arguments, that it factorizes as $J^{S}_{i\alpha} = u_{i} S_{\alpha}$.

\section{Four-moments Maxwell-Boltzmann collinear closure} \label{app_MB_4}
In this Appendix, we provide a proof of the relations
\eqref{begin_mb_4}-\eqref{end_mb_4} between the fluid moments and the Lagrange multipliers in the case of a Maxwell-Boltzmann distributions with four constraints of the moments, in the collinear approximation.

The equilibrium distribution function is given by Eqs. \eqref{statistic} and \eqref{Hp 4m}. We have
\begin{equation}
F^{\textrm{eq}} = \exp \left(-\beta \mathcal{H}' \right) =  \exp \left(-\beta h_{0}' \right) \left[ \cosh \left(-\beta h'_{z} \right) \sigma_{0} + \sigma_{z} \sinh \left( - \beta h'_{z} \right) \right],
\end{equation}
where $h_{0}'$ and $h'_{z}$ are given by Eqs. \eqref{h0_mb 4}-\eqref{hz_mb 4}.
In order to simplify the notation, we introduce the following definitions: $\gamma_{i} = \lambda^{J}_{iz}$ and $J^{S}_{iz} = J^{S}_{i}$. We first compute the density
\begin{eqnarray}
n
&=&
2 \int \exp \left( - \beta h_{0}' \right) \cosh \left( - \beta  h'  \right)d\bm{v} \nonumber \\
&=&
e ^{-\beta \left(\lambda_{0} + \lambda^{S}_{z}\right) } \int e ^{- \frac{\beta m}{2} \left( \bm{v} - \bm{v_{0}} \right)^{2}}  e ^{ - \beta \bm{\gamma} \cdot \bm{v}} d\bm{v}
+ e ^{-\beta \left(\lambda_{0} - \lambda^{S}_{z}\right) } \int e ^{- \frac{\beta m}{2} \left( \bm{v} - \bm{v_{0}} \right)^{2}}  e ^{ \beta \bm{\gamma} \cdot \bm{v}} d\bm{v}. \nonumber
\end{eqnarray}
Let us first define with $I$ the following integral
\begin{equation*}
I^{0}_{\pm}(v_{0_{i}},\gamma_{i}) = \int e ^{- \frac{\beta m}{2} \left( v_{i} - v_{0_{i}} \right)^{2}}  e ^{ \pm \beta \gamma_{i} v_{i}} dv_{i} = \Gamma^{1/3}(T) e ^{\pm \beta \gamma_{i} v_{0_{i}}} e ^{- \beta \gamma_{i}^{2} / 2 m}.
\end{equation*}
Therefore, we have
\begin{eqnarray}
n
&=&
e^{-\beta \left(\lambda_{0} + \lambda^{S}_{z}\right) } I^{0}_{-}(v_{0_{x}},\gamma_{x}) I^{0}_{-}(v_{0_{y}},\gamma_{y}) I^{0}_{-}(v_{0_{z}},\gamma_{z})
+ e ^{-\beta \left(\lambda_{0} - \lambda^{S}_{z}\right) } I^{0}_{+}(v_{0_{x}},\gamma_{x}) I^{0}_{+}(v_{0_{y}},\gamma_{y}) I^{0}_{+}(v_{0_{z}},\gamma_{z})  \nonumber \\
&=&
2\Gamma(T) \exp \left( -\beta \lambda_{0} \right) \exp \left( -\frac{ \beta \bm{\gamma}^{2}}{  2 m} \right) \cosh \left[ \beta \left( \lambda^{S}_{z} + \bm{\gamma} \cdot \bm{v_{0}} \right) \right].
\label{n_annexe}
\end{eqnarray}
The calculation for $S_{z}$ is quite similar, and we obtain
\begin{equation}
S_{z} = \hbar \Gamma(T)  \exp \left( -\beta \lambda_{0} \right) \exp \left( -\frac{ \beta \bm{\gamma}^{2}}{  2 m} \right) \sinh \left[- \beta \left( \lambda^{S}_{z} + \bm{\gamma} \cdot \bm{v_{0}} \right) \right].
\label{s_z_annexe}
\end{equation}

The calculation of $\bm{u}$ is slightly different. Let us compute explicitly the component $u_x$ (the generalization to the other components is then straightforward):
\begin{eqnarray}
u_{x}
&=&
\frac{2}{n} \int v_{x} e \left( - \beta h_{0}' \right) \cosh \left( - \beta | \bm{h'} | \right)d\mathbf{v} \nonumber \\
&=&
\frac{1}{n} \left[ e ^{-\beta \left(\lambda_{0} + \lambda^{S}_{z}\right) } \int v_{x} e ^{- \frac{\beta m}{2} \left( \bm{v} - \bm{v_{0}} \right)^{2}}  e ^{ - \beta \bm{\gamma} \cdot \bm{v}   } d\bm{v}
+ e ^{-\beta \left(\lambda_{0} - \lambda^{S}_{z}\right) } \int v_{x} e ^{- \frac{\beta m}{2} \left( \bm{v} - \bm{v_{0}} \right)^{2}}   e ^{ + \beta \bm{\gamma} \cdot \bm{v}} d\bm{v} \right]. \nonumber
\end{eqnarray}
Defining the following integral
\begin{equation*}
I^{1}_{\pm}(v_{0_{i}},\gamma_{i})
=
\int v_{i} e ^{- \frac{\beta m}{2} \left( v_{i} - v_{0_{i}} \right)^{2}}  e ^{ \pm \beta \gamma_{i} v_{i}} dv_{i} =
\Gamma^{1/3}(T) e ^{\pm \beta \gamma_{i} v_{0_{i}}} e ^{- \beta \gamma_{i}^{2} / 2 m} \left( v_{0_{i}}   \pm \frac{\gamma_{i}}{m} \right),
\end{equation*}
we obtain
\begin{eqnarray}
u_{x}
&=&
\frac{e ^{-\beta \left(\lambda_{0} + \lambda^{S}_{z}\right)}}{n} \left[ I^{1}_{-}(v_{0_{x}},\gamma_{x}) I^{0}_{-}(v_{0_{y}},\gamma_{y}) I^{0}_{-}(v_{0_{z}},\gamma_{z})
+ e ^{2\beta \lambda^{S}_{z} } I^{1}_{+}(v_{0_{x}},\gamma_{x})  I^{0}_{+}(v_{0_{y}},\gamma_{y}) I^{0}_{+}(v_{0_{z}},\gamma_{z}) \right]  \nonumber \\
&=&
v_{0_{x}} - \frac{2 S_{z}}{n \hbar m}  \gamma_{x} .\nonumber
\end{eqnarray}
The generalisation to the other components gives
\begin{equation}
\bm{u} = \bm{v_{0}} + \frac{2 S_{z}}{ n \hbar m} \bm{\gamma}.
\label{u_annexe}
\end{equation}

We finally compute the spin current, again starting from its $x$ component:
\begin{eqnarray}
J^{S}_{x }
&=&
 \hbar \int v_{i} \frac{h'_{\alpha}}{|\bm{h'}|} \exp \left( - \beta h_{0}' \right) \sinh \left( - \beta | \bm{h'} | \right)d\mathbf{v} \nonumber \\
&=&
\frac{\hbar}{2} e ^{-\beta \left(\lambda_{0} + \lambda^{S}_{z}\right) } I^{1}_{-}(v_{0_{x}},\lambda^{J}_{xz}) I^{0}_{-}(v_{0_{y}},\lambda^{J}_{yz}) I^{0}_{-}(v_{0_{z}},\lambda^{J}_{zz})  \nonumber \\
&~&
- \frac{\hbar}{2} e ^{-\beta \left(\lambda_{0} - \lambda^{S}_{z}\right) } I^{1}_{+}(v_{0_{x}},\lambda^{J}_{xz}) I^{0}_{+}(v_{0_{y}},\lambda^{J}_{yz}) I^{0}_{+}(v_{0_{z}},\lambda^{J}_{zz})  \nonumber \\
&=&
v_{0_{x}} S_{z} - \frac{\hbar \gamma_{x}}{2m}n . \nonumber
\end{eqnarray}
The generalisation to the other components gives
\begin{equation}
J^{S}_{i} = v_{0_{i}} S_{z} - \frac{\hbar n}{2m} \gamma_{i}.
\label{Js_annexe}
\end{equation}
Inverting the relations \eqref{n_annexe}-\eqref{Js_annexe}, we obtain
\begin{equation}
\left\{
\begin{array}{lcl}
 \displaystyle \gamma_{i} &=& \displaystyle \frac{2n \hbar m}{\hbar ^{2} n^{2} + 4 S_{z}^{2}} \left( S_{z} u_{i} - J^{S}_{i} \right), \\ \\ \displaystyle
v_{0_{i}} &=& \displaystyle \frac{1}{\hbar ^{2} n^{2} + 4 S_{z}^{2}} \left( \hbar ^{2} n^{2} u_{i} + 4 S_{z} J^{S}_{i} \right),  \\ \\ \displaystyle
e^{ - \beta \lambda_{0} } &=& \displaystyle \frac{e^{\beta \gamma^{2} / 2m}}{\Gamma(T)} \sqrt{\left( \frac{n}{2} \right)^{2} - \left( \frac{S_{z}}{\hbar} \right)^{2}},  \\ \\ \displaystyle
\lambda^{S}_{z} &=& \displaystyle \frac{k_{B}T}{2}   \ln\left( \frac{n - \frac{2|\bm{S}|}{\hbar}}{n + \frac{2|\bm{S}|}{\hbar}} \right)  - \bm{\gamma} \cdot \bm{v_{0}}.
\end{array}
\right.
\label{final_equations_annexe}
\end{equation}

\end{document}